\documentstyle[12pt,epsfig]{article}
 1
\def\as{\alpha_{\rm S}}

\def\citenum#1{{\def\@cite##1##2{##1}\cite{#1}}}
\def\citea#1{\@cite{#1}{}}

\def\as{\alpha_{\rm S}}

\def\b{\beta}
\def\a{\alpha}

\def\g{\gamma}

\def\l{\lambda}

\def\o{\omega}

\def\s{\sigma}

\def\({\left(}
\def\){\right)}

\def\citenum#1{{\def\@cite##1##2{##1}\cite{#1}}}
\def\citea#1{\@cite{#1}{}}

\def\l1vt{\vec{l_{1\perp}}}

\def\bt{b_{\perp}}

\def\bt2{$b^2_t$}

\def\jol1{$J_0(\,l_{1\perp}\,r_{\perp}\,)$}

\def\citea#1{\@cite{#1}{}}

\def\beq{\begin{equation}}
\def\eeq{\end{equation}}
\def\bea{\begin{eqnarray}}
\def\eea{\end{eqnarray}}

\def\eq#1{{Eq.~(\ref{#1})}}

\def\bbbz{{\mathchoice {\hbox{$\sf\textstyle Z\kern-0.4em Z$}}
{\hbox{$\sf\textstyle Z\kern-0.4em Z$}}
{\hbox{$\sf\scriptstyle Z\kern-0.3em Z$}}
{\hbox{$\sf\scriptscriptstyle Z\kern-0.2em Z$}}}}
\def\npb#1#2#3{    {\it Nucl. Phys. }{\bf B#1} (19#2) #3}
\def\plb#1#2#3{    {\it Phys. Lett. }{\bf B#1} (19#2) #3}
\def\prd#1#2#3{    {\it Phys. Rev. }{\bf D#1} (19#2) #3}

\def\zpc#1#2#3{    {\it Z. Phys. }{\bf C#1} (19#2) #3}

\def\sjnp#1#2#3{   {\it Sov. J. Nucl. Phys. }{\bf #1} (19#2) #3}

\relax
\begin{document}
\begin{titlepage}
\noindent
\begin{flushright}
 April 7  1997\\
TAUP 2416/97
\end{flushright}
\begin{center}
{\Large\bf  HARD INCLUSIVE PRODUCTION AND    }   \\[2.4ex]
{\Large \bf  A VIOLATION OF THE FACTORIZATION }\\[2.4ex]
{\Large \bf  THEOREM}\\[11ex]
{\large E.\, G O T S M A N  $^{a)\,1)}$,\,
\footnotetext{$^{1)}$ Email: gotsman@post.tau.ac.il .}
E.\, LEVIN $^{a)\,b)\,2)}$
\footnotetext{$^{2)}$ Email: leving@post.tau.ac.il .}
\,and \,U.\, MAOR $^{a)\,3)}$
\footnotetext{$^{3)}$ Email: maor@post.tau.ac.il .}}\\[1.5ex]

{${}^{a)}$\it School of Physics and Astronomy}\\
{\it Raymond and Beverly Sackler Faculty of Exact Science}\\
{\it Tel Aviv University, Tel Aviv, 69978, ISRAEL}\\[3.5ex]
{${}^{b)}$\it Theory Department, Petersburg Nuclear Physics Institute}\\
{\it 188350, Gatchina, St. Petersburg, RUSSIA}\\[3.5ex]
\end{center}

{\large \bf Abstract:}

 A new mechanism for hard inclusive production, which leads to a violation
of the factorization theorem, is suggested. The mechanism is illustrated 
by a detailed discussion of Higgs meson production in high energy 
deutron-deutron scattering.
 Numerical  
estimates for the effect  are  given for  high
energy hadron (nucleus)  scattering.

\end{titlepage}

\newpage

\par In this letter we consider a new mechanism for hard inclusive
production, which violates the factorization theorem \cite{FACT}.
We illustrate this mechanism for the case of inclusive heavy Higgs production 
in deutron-deutron high energy scattering. The suggested mechanism is
present in
 any central hard inclusive  production, such as the high $p_{T}$
inclusive
production of mini jets and jets.

\par The usual description of inclusive production of a Higgs meson
in a deutron-deutron interaction, is given by a simple Mueller diagram
\cite {MU} shown in Fig.1a.  Our calculation is based on the impulse 
approximation in which the production mechanism is initiated
by a primary single nucleon-nucleon collision.
 Assuming the factorization theorem, the cross section of interest
is given  by  
\beq \label{FT}
\s(Higgs)\,\,=\,\,\int  d x_1 d x_2 F^p_D (x_1,M^2) F^p_D (x_2,M^2)
\,\s(hard)\,\,,
\eeq
where $F^p_D(x,M^2)$ denotes the parton distribution within  the deutron,
 and in the impulse approximation $ F^{p}_{D} = 2F^{p}_{N}$.
$\s(hard)$ denotes the cross section for  Higgs meson production in the
parton - parton collision.

\par Since the deutron is a two nucleon bound state, we may also consider a
production mechanism in which two nucleon-nucleon interactions are responsible
for 
the Higgs production. These are shown in Fig.1b which, after squaring,
gives
the Mueller diagrams of Fig.1c. However, to obtain the  full answer a
more
careful treatment is required
in which we sum over all contributions to the inclusive production,
according to the AGK cutting rules \cite{AGK}. This procedure leads to a
 cancelation of the contributions shown in Fig.1c, so that the net
inclusive
cross section is given by Fig.1a.

\par The above consideration fails to include the interference diagram
shown in Fig.1d. We claim that this contribution survives the
cancelations
implied by the AGK cutting rules, and this results in a violation of the 
factorization theorem. In this note we calculate this contribution and 
assess  the consequences in detail. We wish to remind the reader that
the application of the AGK cutting rules are different when calculating 
inelastic and inclusive processes. This is shown in Fig.2 for
nucleon-nucleon scattering.

\par The present paper continues the approach utilized in many
cascade models for inclusive production in nucleus-nucleus interactions
(see, for example, \cite{CASCADE}). Recently, this mechanism has
been revived by Ryskin \cite{RYS}, who pointed out its importance
for nucleus-nucleus collisions where
its contribution is proportional to $(A_{1}A_{2})^{\frac{4}{3}}$,
wheras using the  factorization theorem  one obtains only an $(A_1 A_2)$
 dependence.
The result obtained by Ryskin is evident from Fig.1d as the contribution
of this diagram is proportional to $ (A_{1}^{2}A_{2}^{2})$ - the
number of two nucleon-nucleon collisions -divided by 
 $\pi R^2_{A_1} \pi R^2_{A_2}$-  which comes from the
integration over  $q_1$ and $q_2$ (the momentum transfers in Fig.1d).
Our paper differs from Refs. \cite{CASCADE} \cite{RYS}, in as much as
we have included explicitly  the relevant contributions and cancelations
implied by the AGK cutting rules. For  a
deutron-deutron collision, we find  that the interference diagram
resulting from the two nucleon interaction (Fig.1d) has to be added to
the 
one nucleon interaction of Fig.1a. Checking the numerics, we obtain a 
significant  addition to the cross section for central mass production,
which
is in the few GeV range. This is relevant for estimates of the  minijet
cross section. 

\par In Fig.3 we show the diagrams illustrating the new mechanism for a 
centrally produced Higgs meson. Fig.3a shows the process where the 
Higgs, which is  produced  centrally, is seperated
 by two large rapidity gaps (LRG) 
  from the small final state multiplicities
which  occur on the edges of the rapidity plot.
We define this process as a double Pomeron exchange reaction, and denote its
contribution as   $\s^{(0)}_2$. The contribution  of Fig.3b is denoted
by $\s^{(1)}_2$. This is a mixed diagram, where the Higgs which is
produced in a double Pomeron process, is superimposed  on a normal
uniform rapidity distribution typical of an inelastic nucleon-nucleon
reaction. Finally, Fig.3c describes Higgs production as part
of the nucleon-nucleon background to the rapidity distribution. We denote
this contribution by $\s^{(2)}_2$. To obtain our final result we need to
sum over all three of the above contributions, noting that these
  are not necessarily positive.

\par The nucleon-nucleon (NN) amplitude for Higgs production via 
 double Pomeron exchange, has been calculated \cite{BL}, to be
\beq \label{HP}
A_H\,=\,A(NN\,\rightarrow\,N + (LRG) + H + (LRG) + N)\,\,=\,\,2 g_H
A_P\,\,,
\eeq
 $g_H$ is the the vertex of the hard parton - parton $\rightarrow $ Higgs
process and  $A_P$ is the Pomeron exchange amplitude.
Fig.3d illustrates graphically the source of the  factor 2
(in \eq{HP}), which plays an important role in our calculation.
The second ingredient in our calculation is the amplitude shown in Fig.3e.
This amplitude has no analog in the case of a single nucleon-nucleon
 interaction,
and it depicts the cut in the diagram shown in  Fig.3b. We note that
this diagram is equal to $Im A_H$, unlike the case for inelastic
 nucleon-nucleon cross section where  $\s_{in} = 2 Im A_P$. The above
result follows  from the unitarity constraint \cite{AGK}.

\par Recalling that
 the  integration over the longitudinal components of the vector $q_{\mu}$ in
Fig.3b results in  a  negative sign for the interference diagram (see Ref.
\cite{AGK} for details),  we can easily calculate 
all diagrams contributing to the inclusive Higgs meson production.
Indeed, calculating $\s^{(0)}_2$,  the sum of  the diagrams in Fig.3a,
and
using
\eq{HP}, one has

\beq \label{S0}
\s^{(0)}_2\,\,=\,\,8 g^2_H \,\{\, (ReA_P)^2\,\,+\,\,(ImA_P)^2\,\}\,\,.
\eeq
For the sum of diagrams in Fig.2b  we have
\beq \label{S1}
\s^{(1)}_2\,\,=\,\,-\,16\,g^2_H\,(Im A_P)^2\,\,.
\eeq
Finally, for $\s^{(2)}_2$, shown in Fig.2c, we obtain
\beq \label{S2}
\s^{(2)}_2\,\,=\,\,8\,g^2_H\,( Im A_p)^2\,\,.
\eeq
The relation between the  different  contributions
is
\beq \label{R012}
\s^{(0)}_2\,:\,\s^{(1)}_2\,:\,\s^{(2)}_2\,\,=\,\,(1+
\rho^2)\,:\,-\,2\,:\,1\,\,,
\eeq
where $\rho = \frac{Re A_P}{Im A_P}$. This  should be compared with the AGK
 relation\cite{AGK}  for the inelastic cross sections
\beq \label{IN}
\s^{(0)}_{in}\,:\,\s^{(1)}_{in}\,:\,\s^{(2)}_{in}\,\,=\,\,(1+
\rho^2)\,:\,-\,4\,:\,2\,\,.
\eeq
Summing all contributions we obtain an additional term in the cross
section compare to the one
given in Eq.(1)
\beq \label{SUM}
\s(D + D \,\rightarrow\,H + X)\,\,=\,\,8\,g^2_H\,(Re A_P)^2\,\,.
\eeq

 We would   like to draw  attention to the novel fact,
 that the contribution of the new
mechanism  is proportional to the real part of the amplitude.
For a soft Pomeron with $\alpha_{P}(0)\; =\; 1$, the contribution of Eq.(8)
vanishes, and we recover the factorization theorem \cite{FACT}, i.e.
an exact cancellation of the diagrams shown in Fig.1c. 

\par  The detailed kinematics of a double Pomeron  Higgs meson production,
 as well as its
amplitude, have been  discussed in Ref.\cite{BL}. The most important
 kinematical feature  is  that at high energy and for
a  Higgs meson in the central rapidity region: y$\,\gg$ 1, and 
Y - y $\gg$ 1 ( see Fig.4a). Accordingly,  the  momentum transfers 
 $q_1$ and $q_2$ are
a small fraction
  of the longitudinal components, 
$\frac{\mid q_{1}\mid}{\mid p_{2} \mid}$  and
$ \frac{\mid q_{2}\mid}{\mid p_{1} \mid} \;\; \ll \;\; 1$.
 We can, therefore, write
$q^2_1\,=\,q^2_{1t}$ and $q^2_2\,=\,q^2_{2t}$.  Expanding
 these momenta with respect to $p_1$ and $p_2$ we have
\beq \label{M}
q_{1\mu}\,\,=\,\,x_1 p_{1\mu}
\,\,+\,\,\b_1\,p_{2\mu}\,\,+\,\,q_{1t;\mu}\,\,,
\eeq
$$
q_{2\mu}\,\,=\,\,x_2 p_{2\mu}
\,\,+\,\,\a_2\,p_{1\mu}\,\,+\,\,q_{2t;\mu}\,\,.
$$
We  obtain
\beq \label{KIN}
x_1\,\,=\,\,\frac{\sqrt{M^2 + (q_1 -q_2)^2_t}}{\sqrt{s}}\,\,e^{y}\,\,\gg\,\,\a_2\,\, ,
\eeq
$$
x_2\,\,=\,\,\frac{\sqrt{M^2 + (q_1
-q_2)^2_t}}{\sqrt{s}}\,\,e^{-y}\,\,\gg\,\,\b_2\,\, ,
$$
where M denotes the Higgs mass.

\par We  write the expression for the diagram of Fig.4a  assuming 
that $k_t\,\,\gg\,\,q_{it}$ ( see Fig.4a). Indeed, $q_{1t}\,\sim\,q_{2t}
\,\approx \,\,\frac{1}{R}$, where R is the hadron (nucleon or nucleus)
radius, while $k_t$ can be as large as M. 
Inserting  the vertex for Higgs emission by a gluon - gluon
fusion \cite{HIGGS}, namely $\Gamma_H\,\,=\,\,g_H ( k_{1\mu} k_{2 \nu} -
g_{\mu \nu} k_1.k_2 ) $, we  obtain the following formula for the Higgs
meson production amplitude of Fig.4a
\beq \label{FOR}
A_H\,\,=\,\,\frac{1}{2} g_H \int d k^2_t \phi(x_1,k^2_t)
\,\,\phi(x_2,k^2_t)\,F_1(q^2_{1t}) F_2(q^2_{2t})\,\,,
\eeq
where $\as(k^2) xG(x,k^2) \,\,=\,\,\int^{k^2} d k'^2 \as(k'^2)
\phi(x,k'^2)$,  $F_1$ and $F_2$ are the form factors describing the $t$
dependence of the Pomeron - target vertex ($F_1(0) = F_2(0) = 1$) .
We wish  to stress that $\phi(x,k^2)$  are not the usual
gluon densities, but the  asymmetric (off diagonal) density
functions that have been introduced in Ref.\cite{GLR}, and have been
discussed in more detail in Ref.\cite{AF}. 
In the leading log(1/x) approximation of perturbative QCD \cite{AF},
 they can be approximated
 in the region of very small $x_1$ and $x_2$
by  the usual nucleon  gluon densities.

Neglecting the $k^2$ dependence of $\as(k^2)$,  $\phi \,=\,\frac{d
xG(x,k^2)}{d k^2}\,\propto (k^2)^{<\gamma> - 1}$, where $<\gamma>$, the
average anomalous dimension,  can be considered to be  a smooth
function of both $x$ and $k^2$ in the region of small $x$. This is a
semi-classical approximation, which holds at small $x$ \cite{GLR}, and we
can use it to estimate  the integral in question. Substituting  
$\phi$ in \eq{FOR} we see that for all $<\gamma>\,<\,\frac{1}{2}$ the
integral is infrared divergent. Only large distances contribute to the
integral and  we can, therefore, replace this integral by the exchange of a
soft Pomeron as was done in Ref.\cite{BL}. In most of the  models for the
soft Pomeron \cite{SP}, the real part of the Pomeron amplitude is very
small, therefore, we expect that  the new mechanism  will only have a very
small contribution 
 to  soft inclusive production.

\par In general,  $<\g>$ is  a function of $x$ and $k^2$,
  approaching  the
value $<\g> = \frac{1}{2}$ in some kinematic region. In the vicinity of
$<\g> = \frac{1}{2}$  the integral over $k^2$ becomes logarithmic and all
distances ($ r_{\perp}$),  down to very small  $
r_{\perp}\,\approx\,\frac{1}{M}$, contribute to the integral. In this case
we need to consider two new physical phenomena: the BFKL Pomeron
 \cite{BFKL} and
 shadowing corrections (SC) \cite{GLR} \cite{12}.
 For the BFKL Pomeron, the value
of the anomalous dimension cannot be larger  than $\g = \frac{1}{2}$ so
we
evaluate  our integral considering $\g \rightarrow \frac{1}{2}$. In this
 case 
\beq \label{BFKL}
Re A_P\,\,=\,\,\frac{\pi}{2} \frac{d Im A_P}{d \ln(1/x)}\,=\,\frac{\pi}{2}
\omega_L Im A_P\,=\,2 \as N_c \ln 2\,\,,
\eeq
where $\alpha_P(0) = 1 + \o_P$ and we have used the explicit calculation
of $\o_P\,=\,\o_L$
\cite{BFKL}. One can
see that for $\alpha_{P}(0) = 1.25$, $Re A_P \,\approx \,Im A_P$. Therefore, we
expect a considerable violation of the factorization theorem for the BFKL
Pomeron, as well as for a DGLAP Pomeron with an intercept which is
well above unity
\cite{GRV}.

\par An estimate  of  SC in the case of Higgs
production via double soft Pomeron exchange in nucleon-nucleon scattering
 has been discussed in \cite{12}.
For the DGLAP evolution    SC become important  in the vicinity of $\g
=
\frac{1}{2}$, where  $Im A_P \,\propto
(\frac{1}{x})^{\o_{cr}}$ \cite{GLR} with $\o_{cr}\,=\,\frac{2 N_c
\as}{\pi}$. Once again, for $\alpha_{P}(0) $= 1.25, we have
 $Re A_P \,\approx Im
A_P$.

 We can, thus,  expect  a considerable violation of the
factorization theorem for  hard inclusive production,  in the kinematic
region where the anomalous dimension approaches $\frac{1}{2}$ and  where
$\o_p$ is well above 0. 

To give a numerical estimate for the additional cross section we write 
\beq \label{CS}
\frac{d\s^{NF}}{d y}\,=\,\int \frac{d q^2_{1t} d q^2_{2t}}{(16 \pi)^2}
|A_H|^2\,\,=\,\,
\frac{2 \pi^2}{ R^2_1 R^2_2} \,\, \s (hard)
 | \int^{M^2}_{Q^2_0} d k^2_t \frac{\partial \phi_1(x_1,k^2)}
{\partial \ln (1 /x_1)}\,\phi_2(x_2,k^2)|^2\,\,,
\eeq
where $\s(hard)$ denotes the hard cross section, which is the same as in
the
factorization theorem (see \eq{FT}), $R_1$ and $R_2$ are the radii of the
colliding hadrons (nuclei) normalized so that: $F_i (t)\,=\,
1 + \frac{1}{4} R^2_i t $. For the nucleus - nucleus collision $\phi_i =
A_i \phi_N$ and $R^2_i = r^2_0 A^{\frac{2}{3}}_i$ and we get $\s^{NF} \,
\propto \,A^{\frac{4}{3}}_1\,A^{\frac{4}{3}}_2$.

\par In Fig.5 we plot the ratio of the nonfactorized contribution of
\eq{CS}  
 to the cross section from the factorization theorem of \eq{FT}, namely
\beq \label{R}
R\,\,=\,\,\frac{2 \pi^2}{R^2_1 R^2_2}\frac{
 | \int^{\frac{M^2}{4}}_{Q^2_0} d k^2_t \frac{\partial \phi_1(x_1,k^2)}
{\partial \ln (1 /x_1)}\,\phi_2(x_2,k^2)|^2}{ x_1 G_1(x_1,\frac{M^2}{4})
x_2 G_2(x_2,\frac{M^2}{4})}\,\,,
\eeq
where $x_i G_i (x_i,M^2)$ is the gluon density  of the $i$-th
hadron. In Fig.5 we took $y$ = 0. 
 Notice, that this ratio does not depend
on the hard cross section, and as such is also applicable to any
central
production process, in particular, hard minijet and jet production.
In evaluating
\eq{R}  
we use the GRV parametrization
\cite{GRV}  
 for the gluon density.
We assume that  $R^2_1 =
R^2_2  = 5\,GeV^{-2}$ (see Ref.\cite{GLMSLOPE} ).
For the initial virtuality we take $Q^2_0 = 1\, GeV^2$, as  the GRV
 parametrization
is in agreement with
  the HERA data
on $F_2(x,Q^2)$ \cite{HERA}   
 for all $Q^2
\,\geq\,1\,GeV^2$.

One can see, that $R$ is quite big for low masses and decreases
when M increases.  For example, $R$ = 1.2   for $M = 10\; GeV$.
 We do not expect
the  Higgs meson  mass to be that small, but our result is also 
applicable in
the case of
  jet production (see Fig.4b)  where   $M \approx 2
p_{\perp}$.  Accordingly, with  such a value of $R$ we can expect  minijet
production (jets with $p_{\perp} \approx 5 GeV$ ) which can be responsible
for the structure of the minimum bias events at the Tevatron energy.

The value of $R$ is bigger for  nucleus - nucleus interaction.
 To estimate
this value we need to multiply \eq{R} by a factor of  $A^{\frac{1}{3}}_{1 eff}
\,A^{\frac{1}{3}}_{2 eff}$. Using the simple relation 
$R^2_A\,=\,r^1_0 A^{\frac{2}{3}}$, one has
$A^{\frac{1}{3}}_{eff}\,=\,\frac{R^2_1}{r^2_0} (\,A^{\frac{1}{3}} 
\,-\,1\,)
\,+\,1$. Therefore, for a  gold - gold interaction we  expect that
 $R$ is enhanced  by factor 4 - 9.

In calculating $R$ we have not included a Sudakov form factor, which is
believed to be necessary 
 in a diagram of the type of Fig.5a, i.e. there is no emission of 
gluons with rapidities between $y_1$ and $y_2$ in Fig.4 (see
Ref.\cite{KMR} for details). However, as  we have
calculated an inclusive production process, this type of emission  should
 be included  (see below). As a first approximation
we calculate the ratio $R$ without the Sudakov form factor
and  neglect the BFKL enhancement, which  can be important for large values of
M (see Fig.6). 

  The diagrams of Fig.4  are only a low order
approximation for the new process. Many  additional gluons
can be emitted, which lead to more complicated  diagrams, shown in
Fig.6.  The emission of the additional gluons can be
summed, and the cross section can be described by a Mueller diagram with
the Pomeron - Pomeron interaction ($\lambda$) as seen in Fig.6. 
It is important to note that we get the same equation (see \eq{SUM})
for the  inclusive Higgs cross section, when using
 the AGK cutting rules
for the Pomeron - Pomeron amplitude
($\lambda$) as was  obtained in Ref.\cite{BR}. Indeed, it was shown  in
Ref.\cite{BR} that for different cuts of $\lambda$ we have a transition
matrix which can be written in the following form
\beq \label{LAMBDA}
\lambda^2_0\,:\,\lambda^1_1\,:\,\lambda^0_2\,:\lambda^2_2\,\,=\,\,
1\,:\,1\,:\,\frac{1}{2}\,:\,\frac{1}{2}\,\,\,
\eeq
with all other  $ \lambda^f_i$ are equal to zero. Here, we denote by $
\lambda^f_i$ the elements of the transition matrix for the Pomeron -
Pomeron amplitude, where $i$ ($f$) is the initial (final)  multiplicity
state (say,
above (below)  the Pomeron - Pomeron interaction in Fig.6) and $i$
($f$) can be equal to 0,1 and 2, respectively. Using this transition matrix
together with the AGK cutting rules for the  inelastic cross section,
 one  obtains the  inclusive cross section
for  Higgs  production
\beq \label{GEN}
\frac{d \s}{d y} (D\,+\,D \,\rightarrow \, H + X)\,\,
\eeq
$$
=\,\,\rho^2 \,\frac{d
\s}{d y} (D\,+\,D \,\rightarrow\, M_1 (n \approx 1)\, + \,(LRG)\,+\,
( H \,+\,X)\,+\, (LRG) \,+\,M_2 (n \approx 1))\,\,,
$$
where $M_1$ and $M_2$ denote the  bunches of final particles with low
 multiplicities. \eq{GEN} provides a method for
calculating the inclusive Higgs  production in terms of  the cross section
for the process with two large rapidity gaps, and the Higgs meson in the
central rapidity region  accompanied by a bunch of final hadrons.

\par  The main result of our paper is \eq{R}, which gives the scale for
the violation of the factorization theorem due to the new mechanism
present in  the
inclusive production. It is clear, that this formula describes not only
the process of  a heavy Higgs meson production, but  also other 
 central hard
processes. The  only difference is   in the explicit expression for
the hard
cross section in  \eq{CS}. From \eq{R} one can see that the new mechanism
only gives  a small contribution to the inclusive cross section for the
production of light hadrons with small transverse momemtum (soft
 inclusive production), because
the
real part of the elastic amplitude is very small at high energy. 
For  hard inclusive production, the new mechanism yields  a sizeable
 contribution, especially in the kinematic region where the anomalous
dimension of the gluon density is close to $\frac{1}{2}$.

Our general conclusion is that the factorization theorem\cite{FACT} is
only approximate.
 We suggest that its  proof should
be reexamined  with the view of finding the  source of the  violation.

\section*{Figure Captions}
\begin{tabular}{l l}
 & \\
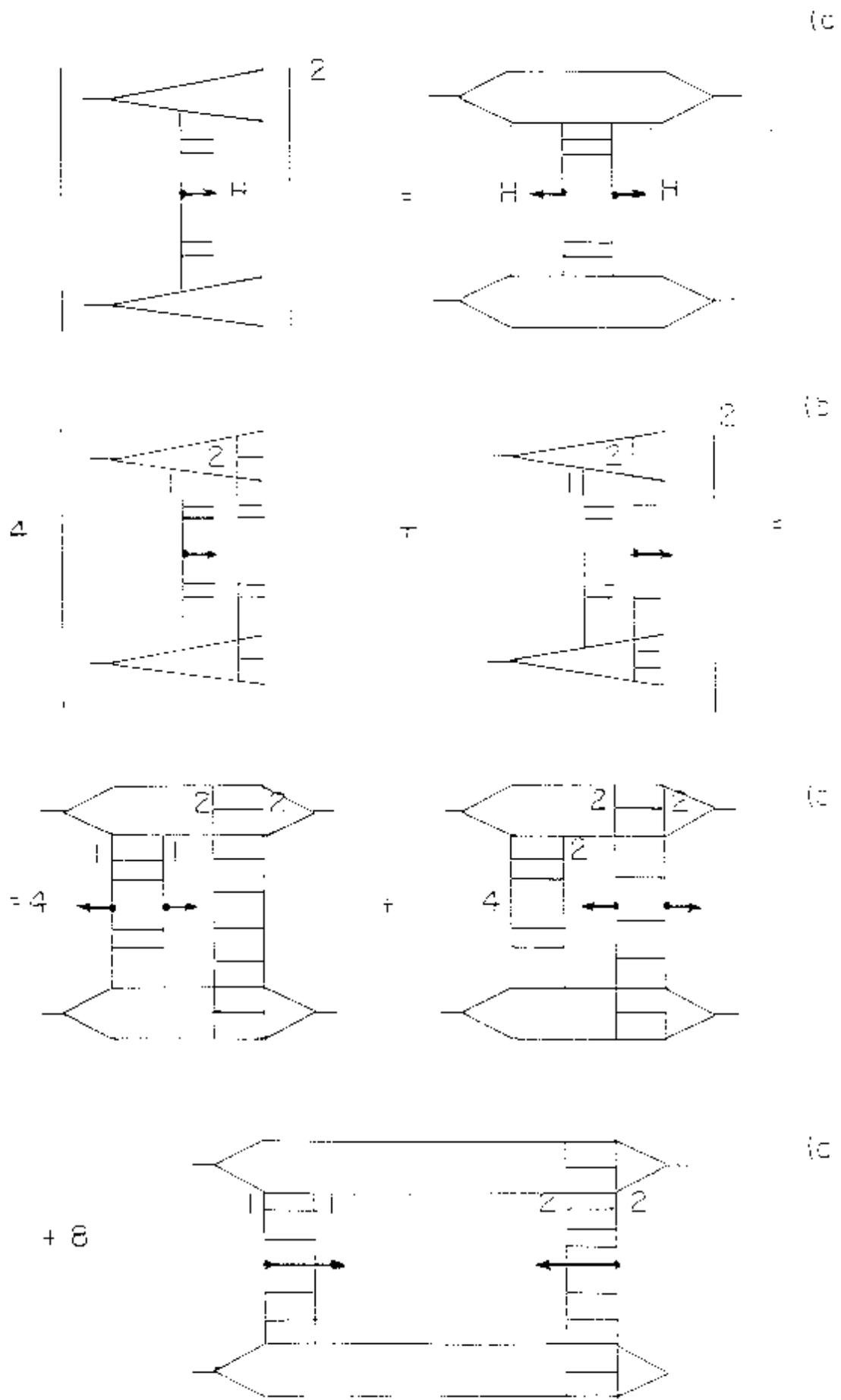
{\bf Fig.1:} & Mueller diagrams for inclusive Higgs production in a 
deutron
- deutron collision :\\
 &  (a) the
factorization theorem contribution, (b)  the two nucleon
interactions,\\
 &(c)  the diagrams, that cancel in total inclusive cross section, (d)
the
interference\\
 & diagram that should be addded to factorization theorem
contribution.\\
 & \\
{\bf Fig.2:} & The AGK cutting rules for inelastic  and inclusive nucleon
- nucleon cross sections. \\
 & \\
{\bf Fig.3:} & Mueller diagrams for the new mechanism of the inclusive
production:\\
& (a)  $\s^{(0)}_2$, (b) $\s^{(1)}_2$, (c) $\s^{(2)}_2$
 and the
relation between different\\
&  Mueller amplitudes ((d)  and (e) ).\\
 & \\
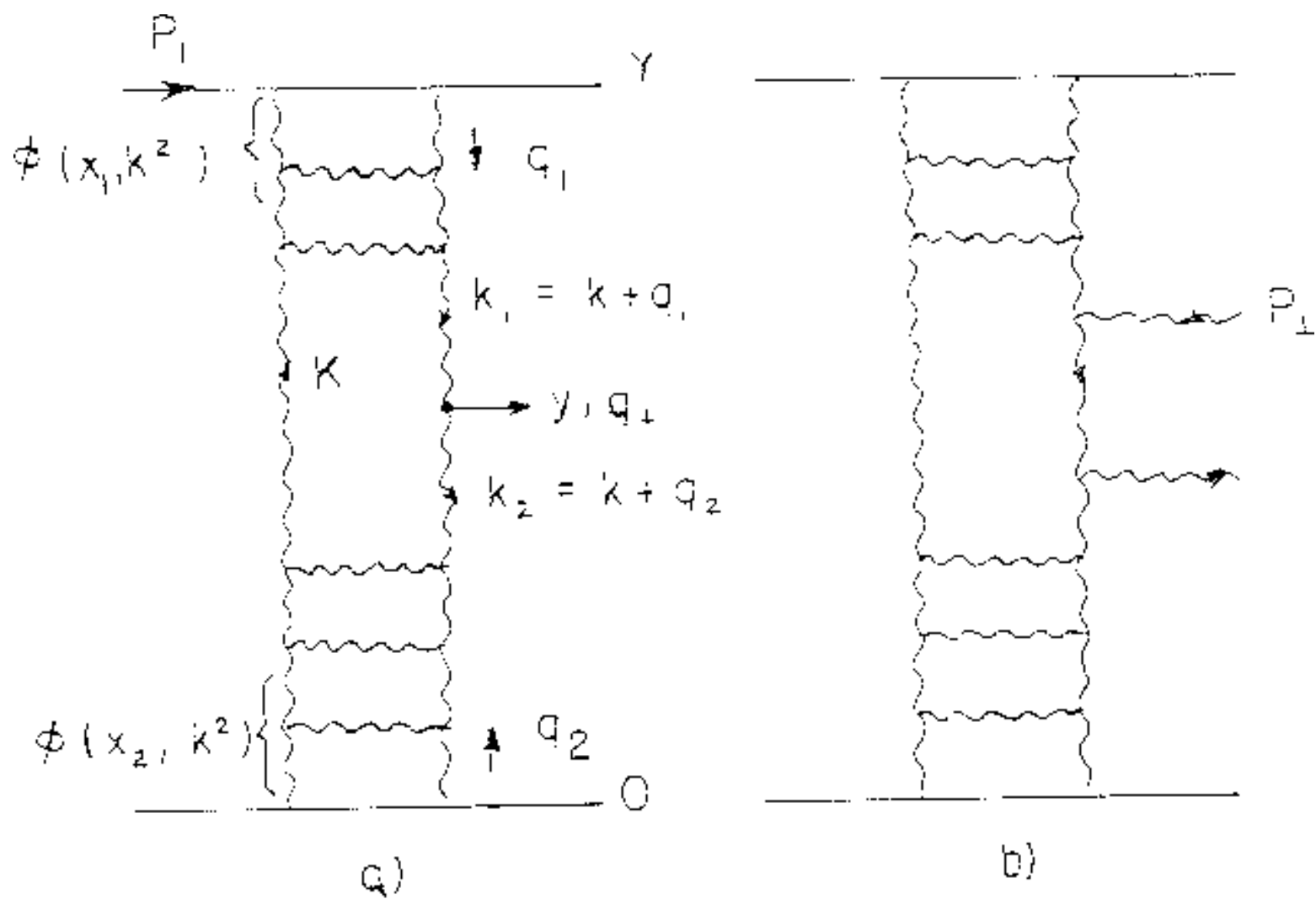
{\bf Fig.4:}&  Feyman diagrams  (a) for Higgs  and (b) for two high $p_T$ 
 jet  production.\\
 &\\
{\bf Fig.5:} & Calculated $R$ (see Eq.(14)).\\
&\\
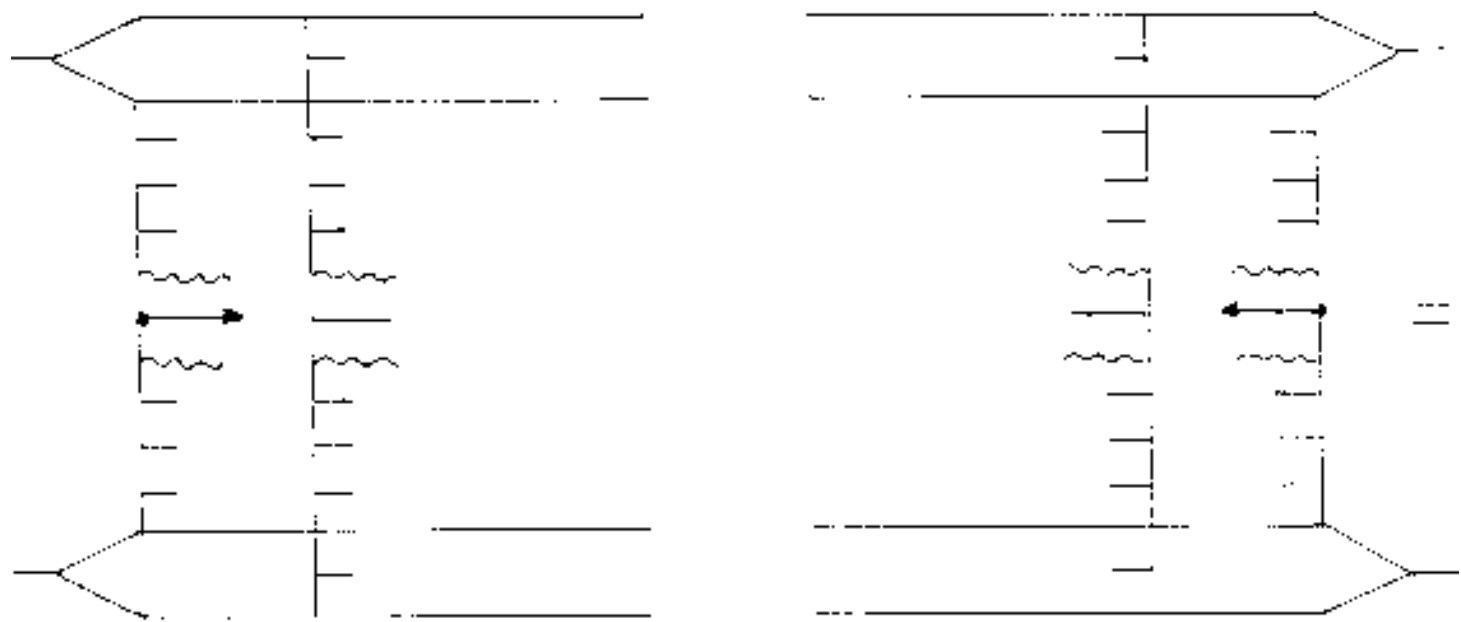
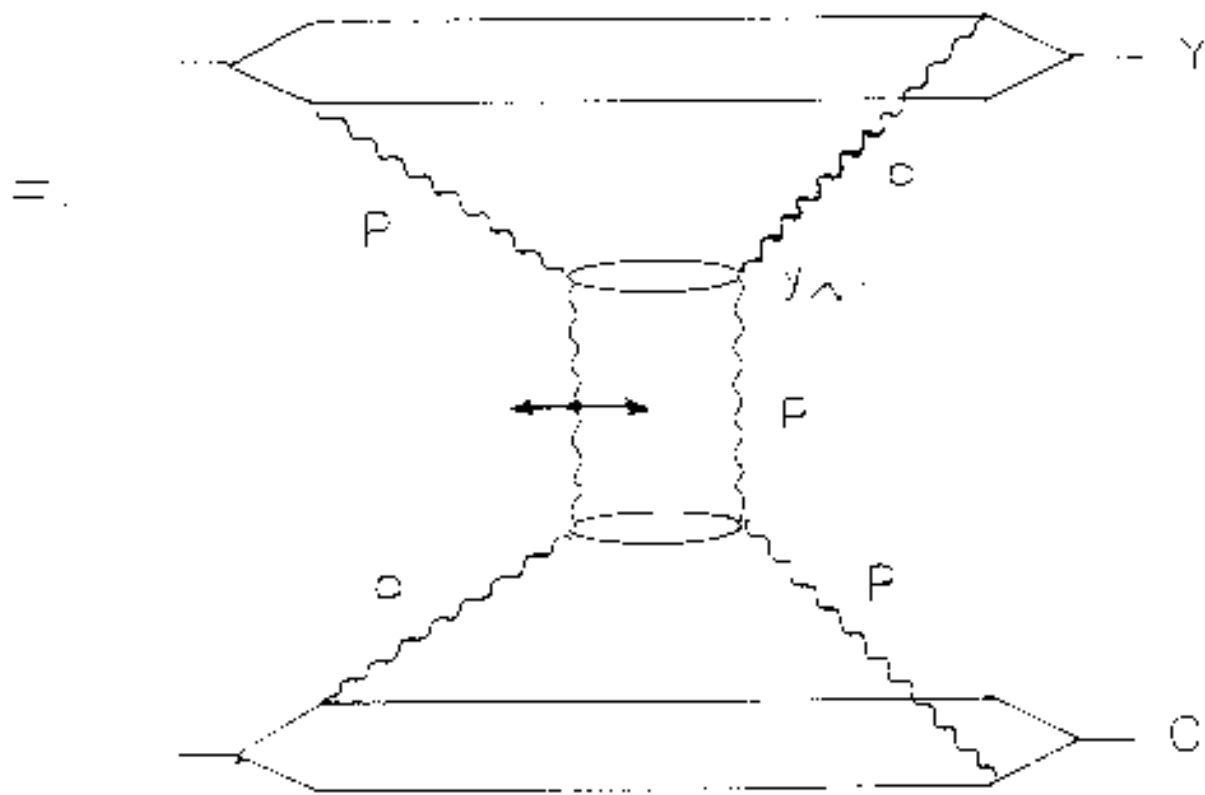
{\bf Fig.6:} & The general Mueller diagram for the new mechanism\\
 &  for  the
hard inclusive production.\\
\end{tabular}

   
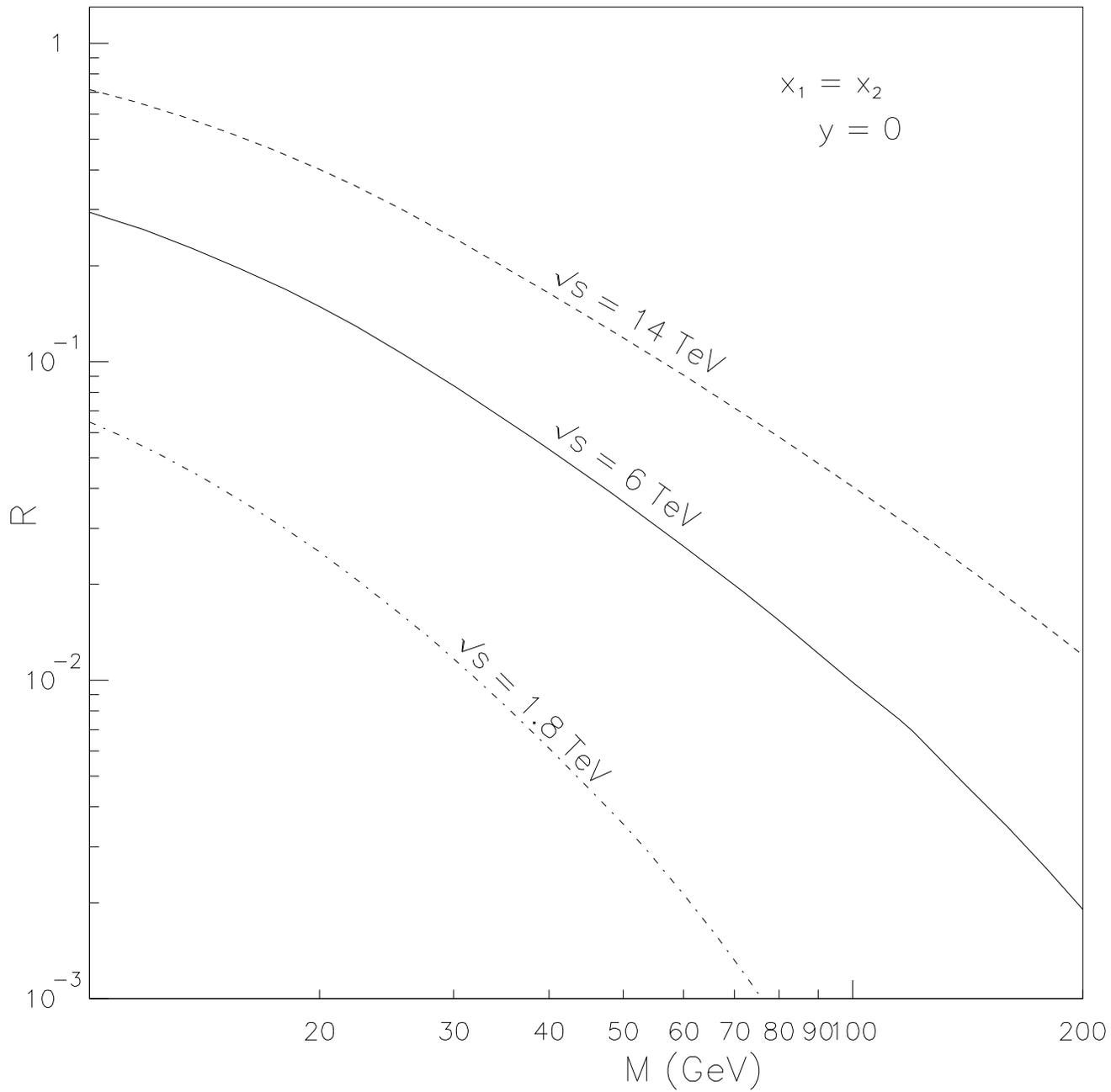
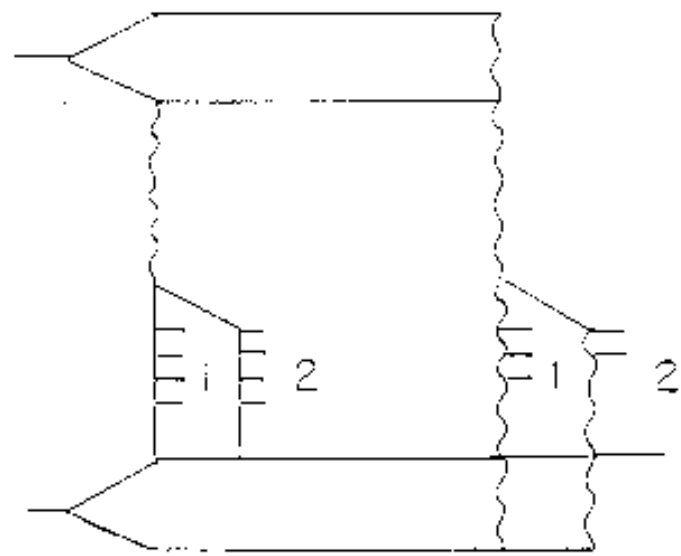
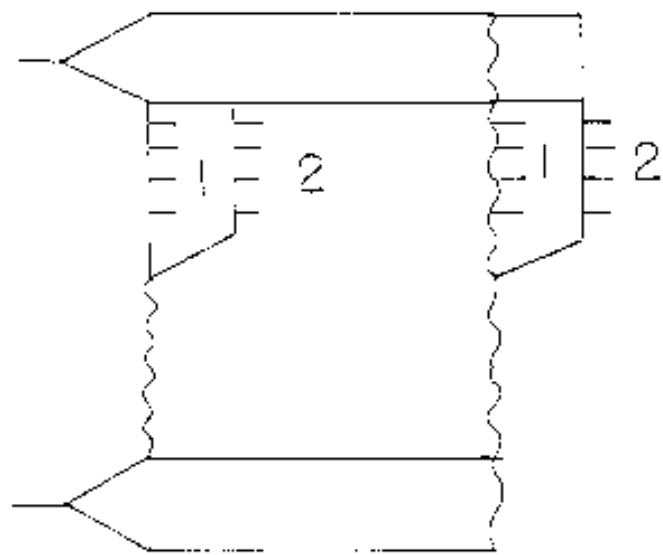
\end{document}